\documentclass[12pt]{article}
\usepackage{amssymb}

\def\1{{1\mskip-10mu1}}

\def\bea{\begin{eqnarray*}}
\def\eea{\end{eqnarray*}}
\def\bean{\begin{eqnarray}}
\def\eean{\end{eqnarray}}

\begin{document}

\title{\bf On Quantum - Classical Correspondence 
 for
Baker's Map}
\author{Kei Inoue$\dagger $ , 
Masanori Ohya$\dagger $ and Igor V. Volovich$\ddagger $
\\~~\\
$\dagger $Department of Information Sciences\\
Science University of Tokyo\\
Noda City, Chiba 278-8510 Japan\\~\\
$\ddagger ~$Steklov Mathematical Institute\\
Russian Academy of Sciences\\
Gubkin St. 8, 117966 Moscow, Russia\\
{\it e-mail: volovich@mi.ras.ru}
}
\date{}
\maketitle

\begin{abstract}
Quantum baker`s map is 
a model of chaotic system.
We study quantum dynamics  for the quantum baker's
map. We use the Schack and Caves symbolic
description of the quantum baker`s map.
We find  an exact expression 
for the expectation value of the time dependent position operator.
 A relation between quantum  and  classical 
trajectories is investigated. Breakdown of the quantum-classical
correspondence 
at the logarithmic timescale  is rigorously
established. 
\end{abstract}

\newpage
\section{Introduction}

The quantum-classical
correspondence for dynamical systems has been studied for many years, see
for example \cite{Zur1,GJK} and reference therein. 
A significant progress
in understanding of this correspondence has been achieved in a
mathematical
approach when one considers the Planck constant $h$ as a small variable
parameter. It is well known that in the limit $h\rightarrow 0$ quantum
theory is reduced to the classical one \cite{Mas,Hep}. 

However in physics the Planck
constant is a fixed constant although it is very small. Therefore it is
important to study the relation between classical and quantum evolutions
when the Planck constant is fixed. There is a conjecture 
\cite{BZ,BB,ZP} that
a characteristic timescale $t_h $ appears in the quantum evolution of
chaotic dynamical systems. For time less then $t_h $ there is a
correspondence between quantum and classical expectation values, while for
times greater that $t_h$ the predictions of the classical and quantum
dynamics no longer coincide.

 An important problem is to estimate the
dependence $t_h $ on the Planck constant $h.$ Probably a universal formula
expressing $t_h $ in terms of $h$ does not exist and every model should be
studied case by case. It is expected that certain quantum and classical
expectation values diverge on a timescale inversely proportional to some
power of $h$ . Other authors suggest that for chaotic
systems a breakdown may be
anticipated on a much smaller logarithmic timescale (see \cite
{Zur1,KZZ} for a discussion). Numerous works are devoted
to the analytical and numerical study of 
classical and quantum chaotic systems
\cite{AA} - \cite{IOV}.

 Most results concerning various timescales
are obtained numerically.
In this paper we will present some {\it exact} results on a quantum chaos
model. We compute explicitly an expectation value 
for the quantum baker`s map and prove 
rigorously the appearance
of the logarithmic timescale.

The quantum baker's map is a model 
invented to study the  chaotic behavior \cite{BV}. 
The model has been studied
in \cite{OS} - \cite{SS3}.

In this paper quantum dynamics of 
the position operator for the quantum baker's
map is considered. We use a simple  symbolic
description of the quantum baker`s map 
proposed by Schack and Caves \cite{SC3}.
We find  an exact expression 
for the expectation value of the time dependent position operator.
In this sense the quantum baker`s map is an exactly solvable model
though stochastic one.
 A relation between quantum  and  classical 
trajectories is investigated. For some matrix elements
the breakdown of the quantum-classical
correspondence  
at the logarithmic timescale  is
established. 

Here we would like to note that in fact the notion of the timescale
is not  a uniquely defined notion. Actually we will obtain the formula
$$
\left\langle \hat{q}_m \right\rangle -q_m = h2^{m-1}
$$
where $\hat{q}_m$ and $q_m$ are quantum and classical positions
respectively at time $m$. This formula will be interpreted
as the derivation of the logarithmic timescale (see discussion in Sect.5).
The main result of the paper is presented in Theorem 1 in Sect. 4.

  In another paper \cite{IOV}, semiclassical properties and
chaos degree for the quantum baker's map are considered. 

\section{Classical Baker's Transformation}

The classical baker's transformation maps the unit square 
$0\leq $ $q,p\leq1 $ onto itself according to 
\[
\left( q,p\right) \rightarrow \left\{ 
\begin{array}{ll}
\left( 2q,p/2\right) , & \mbox{if}\quad 0\leq q\leq 1/2 \\ 
\left( 2q-1,\left( p+1\right) /2\right) , & \mbox{if\quad }1/2<q\leq 1
\end{array}
\right. 
\]

This corresponds to compressing the unit square in the $p$ direction and
stretching it in the $q$ direction, 
while preserving the area, then cutting
it vertically and stacking the right part on top of the left part.

The classical baker's map 
has a simple description in terms of its symbolic
dynamics \cite{AY}. 
Each point $\left( q,p\right) $ is represented by a
symbolic string with a dot

\begin{equation}
\xi=\cdots \xi_{\_2}\xi_{\_1}\xi_{0}.\xi_{1}\xi_{2}\cdots ,  \label{2.1}
\end{equation}

\noindent where $\xi_{k}\in \left\{ 0,1\right\} $, and

\[
q=\sum_{k=1}^{\infty }\xi_{k}2^{-k},\qquad p=\sum_{k=0}^{\infty
}\xi_{-k}2^{-k-1} 
\]

\noindent The action of the baker's map on a symbolic string $\xi$ 
is given by
the shift map (Bernoulli shift ) 
$U$ defined by $U\xi=\xi^{^{\prime }}$, where $%
\xi_{m}^{^{\prime }}=\xi_{m+1}$. 
This means that, at each time step, the dot is
shifted one place to the right while entire string remains fixed.
After $m$ steps the $q$ coordinate becomes
\begin{equation}
q_m=\sum_{k=1}^{\infty}\xi_{m+k}2^{-k}  \label{2.1b}
\end{equation}
This relation defines the classical trajectory with
the initial data
\begin{equation}
q=q_0=\sum_{k=1}^{\infty}\xi_{k}2^{-k}  \label{2.1c}
\end{equation}

\section{Quantum Baker's Map}

Quantum baker's maps are defined on the $D$-dimensional 
Hilbert space of the
quantized unit square. To quantize the unite square one defines
the Weyl unitary displacement 
operators $\hat{U}$ and $\hat{V}$ in $D$ -
dimensional Hilbert space, 
which produces displacements in the momentum and
position directions, respectively, 
and the following commutation relation is
obeyed
\[
\hat{U}\hat{V}=\epsilon \hat{V}\hat{U}, 
\]
\noindent where $\epsilon =\exp \left( 2\pi i/D\right) .$ 
We choose $D=2^{N}$, 
so that our Hilbert space will be the $N$ qubit space
$\mathbb{C}^{\otimes N}$. The constant $h=1/D=2^{-N}$ can be regarded 
as the Planck constant.
The space $\mathbb{C}^2$ has a basis
$$
\left| 0\right\rangle =\left(
    \begin{array}{c}0\\1
    \end{array}
    \right),~~
\left| 1\right\rangle =\left(
    \begin{array}{c}1\\
    0\end{array}
    \right)
$$	
The basis in $\mathbb{C}^{\otimes N}$ is
\[
\left|\xi_{1}\right\rangle \otimes \left|
\xi_{2}\right\rangle \otimes \cdots 
\otimes \left| \xi_{N}\right\rangle ,
~~\xi_k=0,1 
\]
We write
\[
\xi=\sum_{k=1}^N \xi_k2^{N-k}
\]
then $\xi = 0,1,...,2^N-1$ and denote
$$
\left| \xi\right\rangle =\left| \xi _{1}\xi _{2}\cdots \xi
_{N}\right\rangle =	
\left|\xi_{1}\right\rangle \otimes \left|
\xi_{2}\right\rangle \otimes \cdots 
\otimes \left| \xi_{N}\right\rangle 
$$	
We will use for this basis also notations $\{\left| \eta
\right\rangle =
\left| \eta _{1}\eta _{2}\cdots 
\eta_{N}\right\rangle,~~\eta_k=0,1\}$
and $\{\left|
j\right\rangle =\left| j _{1}j _{2}\cdots j
_{N}\right\rangle,~~j_k=0,1\}$.

The operators $\hat{U}$ and $\hat{V}$ can be written as
\[
\hat{U}=e^{2\pi i\hat{q}},\quad \hat{V}=e^{2\pi i\hat{p}} 
\]
where the position and momentum operators $\hat{q}$ and 
$\hat{p}$ are operators in $\mathbb{C}^{\otimes N}$
which are defined as follows. The position operator is
$$
\hat{q}=\sum_{j=0}^{2^{N}-1}q_{j}\left|
j\right\rangle \left\langle j\right|=\sum_{j_1,...,j_N}
q_j \left|
j_N...j_1\right\rangle \left\langle j_1...j_N\right|
$$
where
$$
\left|
j\right\rangle =\left| j _{1}j _{2}\cdots j
_{N}\right\rangle,~~j_k=0,1 
$$
is the basis in $\mathbb{C}^{\otimes N}$,
$$
j=\sum_{k=1}^{N}j_{k}2^{N-k}
$$ 
and
$$
q_{j}=\frac{j+1/2}{
2^{N}},~~j=0,1,\ldots ,2^{N}-1
$$
The momentum operator is defined as
$$
\hat{p}=F_{N}\hat{q}F_{N}^*
$$
where $F_{N}$ is the quantum 
Fourier transform acting to the basis vectors as
$$
F_{N}\left|
j\right\rangle=\frac{1}{\sqrt{D}}
\sum_{\xi=0}^{D-1}e^{2\pi i\xi j/D}\left|\xi\right\rangle, 
$$
here $D=2^N$.

A quantum baker`s map is the unitary operator $T$ in 
$\mathbb{C}^{\otimes N}$ with the following matrix elements
\begin{equation}
\left\langle \xi \right| T\left| \eta\right\rangle =
\frac{1-i}{2}\exp \left( 
\frac{\pi }{2}i\left| \xi _{1}-\eta _{N}\right| \right)
\prod_{k=2}^{N}\delta \left( \xi _{k}
-\eta _{k-1}\right)  ,  
\label{4.2}
\end{equation}

\noindent where
$\left| \xi \right\rangle =\left| \xi _{1}\xi
_{2}\cdots \xi _{N}\right\rangle $,~~ 
$\left| \eta \right\rangle
=\left| \eta _{1}\eta _{2}\cdots \eta _{N}\right\rangle $ 
and $\delta (x)$ is the Kronecker symbol, $\delta (0)=1;~
\delta (x)=0, x\neq 0.$
This transformation belongs to a family
of quantizations of baker`s map  introduced  by Schack and Caves
\cite{SC3} and studied in \cite{SS,SS3}.

\section{Expectation Value}

We consider the following  mean value 
of the position operator $\hat{q}$
for time $m=0,1,...$ with respect to a  vector $\left| \xi
\right\rangle $ :
\begin{equation}
r_{m}^{\left( N\right) }=\left\langle \xi \right| T^{m}\hat{q}
T^{-m}\left| \xi \right\rangle ,  \label{4.1}
\end{equation}
\noindent where $\left| \xi \right\rangle =
\left| \xi _{1}\xi _{2}\cdots \xi
_{N}\right\rangle $.
First we show that there is an explicit 
formula for  the expectation value
$r_{m}^{\left( N\right) }$. In this sense the quantum baker`s map
is an explicitly solvable model. 
Then we compare the dynamics of the mean value $r_{m}^{\left(
N\right) }$ of position operator $\hat{q}$ 
with that of the classical value $%
q_m$ , Eq.~(\ref{2.1b}). We will establish a logarithmic timescale
for the breakdown of the 
quantum-classical correspondence for the quantum baker`s map.

From Eq.~(\ref{4.2}) one gets for $m=0,1,...,N-1$
\begin{equation}
\left\langle \xi \right| T^{m}\left| \eta \right\rangle =
\left( 
\frac{1-i}{2}\right) ^{m}\left( \prod_{k=1}^{N-m}
\delta \left( \xi_{m+k}-\eta _{k}\right) \right) 
\left( \prod_{l=1}^{m}\exp \left(
\frac{\pi }{2}i\left| \xi _{l}-\eta _{N-m+l}
\right| \right) \right) ,
\label{4.3}
\end{equation}
and for $m=N$
\begin{equation}
\left\langle \xi \right| T^{N}\left| \eta \right\rangle =
\left( 
\frac{1-i}{2}\right) ^{N} 
\left( \prod_{l=1}^{N}\exp \left(
\frac{\pi }{2}i\left| \xi _{l}-\eta _{l}
\right| \right) \right) 
\label{4.3N}
\end{equation}

Using this formula we will prove the following 

{\bf Theorem 1.}
One has the following expression for the expectation 
valule (\ref{4.1}) of
the position operator
\begin{equation}
r_{m}^{\left( N\right) }=\left\langle \xi \right| T^{m}\hat{q}
T^{-m}\left| \xi \right\rangle=
\sum_{k=1}^{N-m}\frac{\xi _{m+k}}{2^{k}}+\frac{1}{
2^{N-m+1}}  \label{4.4}
\end{equation}
\noindent for $0\leq m<N$ . For $m=N $ we have
 \begin{equation}
r_{N}^{\left( N\right) }=\frac{1}{2}
 \label{4.4b}
\end{equation}
\noindent {\bf Proof. }By a direct calculation, we obtain 
\begin{eqnarray*}
r_{m}^{\left( N\right) } &=&\left\langle \xi \right| T^{m}\hat{q}
T^{-m}\left| \xi \right\rangle  \\
&=&\left\langle \xi \right| T^{m}\left( \sum_{j=0}^{2^{N}-1}
\frac{j+1/2}{
2^{N}}\left| j\right\rangle \left\langle j
\right| \right) T^{-m}\left|
\xi \right\rangle  \\
&=&\sum_{j=0}^{2^{N}-1}\frac{j+1/2}{2^{N}}\left\langle \xi \right|
T^{m}\left| j\right\rangle \left\langle j\right| T^{*m}\left| \xi
\right\rangle  \\
&=&\sum_{j=0}^{2^{N}-1}\frac{j+1/2}{2^{N}}
\left| \left\langle \xi \right|
T^{m}\left| j\right\rangle \right| ^{2}.
\end{eqnarray*}

\noindent Using (\ref{4.3}) we write 
\begin{eqnarray*}
r_{m}^{\left( N\right) }& =&
\sum_{j=0}^{2^{N}-1}\frac{j+1/2}{2^{N}}
\left| \left( \frac{1-i}{2}\right)
^{m}\left( \prod_{k=1}^{N-m}
\delta \left( \xi _{m+k}-j_{k}\right) \right)
\left( \prod_{l=1}^{m}\exp \left( \frac{\pi }{2}i
\left| \xi_{l}-j_{N-m+l}\right| \right) \right) \right| ^{2} \\
&=&\sum_{j=0}^{2^{N}-1}\frac{j+1/2}{2^{N}}
\left | \frac{1-i}{2}\right |^{2m}
\left( \prod_{k=1}^{N-m}\delta \left(
\xi _{m+k}-j_{k}\right) \right)  \\
&=&\frac{1}{2^{N+m}}\sum_{j_{1},\cdots j_{N}}\left\{ \left(
\sum_{l=1}^{N}j_{l}2^{N-m}\right) +1/2\right\} \left(
\prod_{k=1}^{N-m}\delta \left( \xi _{m+k}-j_{k}\right) \right)  \\
&=&\frac{1}{2^{N+m}}\sum_{j_{1},\cdots j_{N}}\left(
\sum_{l=1}^{N}j_{l}2^{N-k}\right) \left( \prod_{k=1}^{N-m}
\delta \left( \xi
_{m+k}-j_{k}\right) \right)  \\
&&+\frac{1}{2^{N+m+1}}\sum_{j_{1},\cdots j_{N}}\left(
\prod_{k=1}^{N-m}\delta \left( \xi _{m+k}-j_{k}\right) \right)  \\
\end{eqnarray*}
Using the Kronecker symbols one gets
\begin{eqnarray*}
r_{m}^{\left( N\right) } 
&=&\frac{1}{2^{N+m}}\sum_{j_{N-m+1},\cdots j_{N}}
\left( \sum_{l=1}^{N-m}\xi
_{m+l}2^{N-l}+\sum_{l=N-m+1}^{N}j_{l}2^{N-l}\right) +
\frac{1}{2^{N+m+1}}
\left( \sum_{j_{N-m+1},\cdots j_{N}}1\right)  \\
\end{eqnarray*}
We can write it as
\begin{eqnarray*}
r_{m}^{\left( N\right) } 
&=&\frac{1}{2^{N+m}}\left( \sum_{l=1}^{N-m}
\xi _{m+l}2^{N-l}\right) 
\left(
\sum_{j_{N-m+1},\cdots j_{N}}1\right) +
\frac{1}{2^{N+m}}\sum_{j_{N-m+1},
\cdots j_{N}}\left( \sum_{l=N-m+1}^{N}j_{l}2^{N-l}\right)  \\
&&+\frac{1}{2^{N+m+1}}\left( \sum_{j_{N-m+1},
\cdots j_{N}}1\right)  \\
&=&\frac{2^{m}}{2^{N+m}}\left( \sum_{l=1}^{N-m}
\xi _{m+l}2^{N-l}\right) +
\frac{1}{2^{N+m}}\sum_{j_{N-m+1},\cdots j_{N}}\left(
\sum_{l=N-m+1}^{N}j_{l}2^{N-l}\right) +
\frac{2^{m}}{2^{N+m+1}} \\
&=&\frac{1}{2^{N}}\left( \sum_{l=1}^{N-m}
\xi _{m+l}2^{N-l}\right) +
\frac{1}{%
2^{N+m}}\sum_{j_{N-m+1},\cdots j_{N}}\left(
\sum_{l=1}^{m}j_{N-m+l}2^{m-l}\right) +\frac{1}{2^{N+1}} \\
\end{eqnarray*}
Finally we obtain (\ref{4.4})
for $0\leq m<N$
\begin{eqnarray*}
r_{m}^{\left( N\right) } 
&=&\frac{1}{2^{N}}\left( \sum_{l=1}^{N-m}
\xi _{m+l}2^{N-l}\right) +\frac{1}{
2^{N+m}}\frac{1}{2}\left( 2^{m}-1\right) 
2^{m}+\frac{1}{2^{N+1}} \\
&=&\left( \sum_{k=1}^{N-m}\xi _{m+k}2^{-k}\right) +
\frac{1}{2^{N-m+1}}
\end{eqnarray*}
In the case $m=N$ we have
\begin{eqnarray*}
r_{N}^{\left( N\right) } &=&\sum_{j=0}^{2^{N}-1}
\frac{j+1/2}{2^{N}}\left|
\left\langle \xi \right| T^{N}\left| j\right\rangle \right| ^{2}\\
&=&\sum_{j=0}^{2^{N}-1}\frac{j+1/2}{2^{N}}
\left| \left( \frac{1-i}{2}\right)
^{N} \right| ^{2} 
=\frac{1}{2^{2N}}\sum_{j=0}^{2^{N}-1}(j+1/2)
=\frac{1}{2}.
\end{eqnarray*}
The theorem is proved.
\section{Time Scales}
We consider here the quantum-classical correspondence
for the quantum baker`s map. First let us mention that 
$2^N=1/h$ and the limit $h\to 0$ corresponds to the limit $N\to \infty$.
Therefore from Theorem 1 and Eq.~(\ref{2.1b}) one has the mathematical
correspondence between quantum and classical trajectories as $h\to 0$:
$$
\lim_{N\to \infty} r_{m}^{\left( N\right) }=q_m,~~m=0,1,...
$$

Now let us fix the Planck constant $h=2^{-N}$
and investigate on which time scale the quantum and classical
expectation values start to differ from each other.
From Theorem 1 and Eq.~(\ref{2.1b}) we obtain the following

{\bf Proposition 1.}
Let $r_{m}^{\left( N\right) }$ be the mean value of 
position operator $\hat{q}$ at the time $m$ and $q_m$ is the 
classical trajectory Eq.~(\ref{2.1b}). Then we have 
\begin{equation}
q_{m}-r_{m}^{\left( N\right) }=\sum_{j=N-m+1}^{\infty }\xi _{m+j}2^{-j}-
\frac{1}{2^{N-m+1}}  \label{4.5}
\end{equation}
for any $0\leq m\leq N$.

Let us estimate the difference between the quantum and classical
trajectories.

{\bf Proposition 2.}.
Let $q_{m}$ and $r_{m}^{\left( N\right) }$ be the same as in the 
Proposition 1.
Then we have
\begin{equation}
\left| r_{m}^{\left( N\right) }-q_{m}\right| \leq \frac{1}{2^{N-m+1}}
\label{4.6}
\end{equation}
for any string $\xi=\xi_1\xi_2...$ and any time $0\leq m\leq N$.

\noindent \noindent \noindent {\bf Proof. } Note that
$$
0 \leq \sum_{j=N-m+1}^{\infty }\xi _{m+j}2^{-j} 
$$
$$
\leq \frac{1}{2^{N-m+1}}\left( 1+\frac{1}{2}+\left( \frac{1}{2}\right)
^{2}+\cdots \right) 
=\frac{1}{2^{N-m}}.
$$
\noindent Using the above inequality, one gets from Eq.~(\ref{4.5})
\[
-\frac{1}{2^{N-m+1}}\leq q_{m}-r_{m}^{\left( N\right) }
\leq \frac{1}{2^{N-m}}%
-\frac{1}{2^{N-m+1}}=\frac{1}{2^{N-m+1}}
\]
This means that we have
\[
\left| r_{m}^{\left( N\right) }-q_{m}\right| \leq \frac{1}{2^{N-m+1}}
\]

\noindent for any $0\leq m\leq N$. 

Proposition 2  shows an exact 
correspondence between quantum and classical
expectation value for baker's map.
We can write the relation (\ref{4.6}) in the form 
\begin{equation}
\left| r_{m}^{\left( N\right) }-q_{m}\right|
 \leq \frac{1}{2^{N-m+1}}=h2^{m-1}  \label{4.7}
\end{equation}
since the Planck constant $h=2^{-N}$.
In particular for $m=0$ we have
\begin{equation}
\left| r_{0}^{\left( N\right) }-q_{0}\right| \leq \frac{h}{2}  
\label{4.70}
\end{equation}
for any $\xi=\xi_1\xi_2...$.

Now let us estimate at what time $m=t_h$ there appears an
essential difference
between classical trajectory and quantum expectation value.
From Eq.~(\ref{4.7}) we can expect that the time  $m=t_h$
corresponds to the maximum of the function $2^m/2^{N-1}$ for 
$0\leq m\leq N$, i.e. 
\begin{equation}
t_h=N=\log_2 \frac{1}{h}
\label{4.7h}
\end{equation}
For time $0\leq m<t_h$ the difference
between classical and quantum trajectories in (\ref{4.7}) is 
bounded by $1/4$ since
\[
h2^{m-1}=\frac{1}{2^{N-m+1}}\leq \frac{1}{4}
\]
One can see that the bound is saturated. Indeed let us take
a string $\xi$ with arbitrary $\xi_1,...,\xi_N$ but with
$\xi_{N+1}=0, \xi_{N+2}=0,...$. Then one has
$$
r_{m}^{\left( N\right) }-q_{m}=h2^{m-1},~~m=0,1,...,N
$$
Therefore we have established the logarithmic 
dependence of the timescale on the
Planck constant $h$.

\section{Conclusions}

In this paper we have computed the expectation values for the position
operator in the quantum baker`s map. Breakdown of the quantum-classical
correspondence 
at the logarithmic timescale  is rigorously
established.  For better understanding
of the quantum-classical correspondence and the decoherence
process it is important to 
perform similar computations for more general matrix elements
which include also the momentum operators
 and coherent vectors. 
 
 Only the simplest quantization
 of the baker`s map was considered in the paper.
 It would be interesting to extend the computations to the whole family
 of quantizations of quantum baker`s map proposed in \cite{SC3}.
 Some of these questions will be investigated in another paper
 \cite{IOV}.

\section{Acknowledgements}
We are grateful to R. Schack, Ya.G. Sinai and W. Zurek
for interest to the work and helpful remarks. 
The main part of this work was done during
the visit of I.V. to the Science University of Tokyo.
He is grateful to  JSPS for the  Fellowship award.
His work   was also supported in part by RFFI 99-0100866
and by INTAS 99-00545.

\end{document}